# What matters in the new field of machine learning and satellite imagery-based poverty predictions? A review with relevance for potential downstream applications and development research

(8 000 - 10 000 words)

Ola Hall[a],* Francis Dompae, Ibrahim Wahab[a], Fred Mawunyo Dzanku[b]

[a] *Department of Human Geography, Lund University, Lund, Sweden;* [b]*Institute of Statistical, Social and Economic Research, University of Ghana, Legon, Ghana*

* corresponding author: Ola.hall@keg.lu.se

## Abstract

The field of artificial intelligence – here encompassing Machine Learning, Deep Learning and Transfer Learning – is seeing the increased application of satellite imagery to analyse poverty in its various manifestations. This nascent but rapidly growing intersection of scholarship holds the potential to help us better understand poverty by leveraging big data and recent advances in machine vision. This paper reviews the state-of-the-art in this domain and finds some interesting results. The most important factors correlated to the predictive power of welfare in the reviewed studies are the number of pre-processing steps employed, the number of datasets used, the type of welfare indicator targeted, and the choice of AI model. As expected, studies that used *hard* indicators as targets achieved better performance – 17 percentage points higher – in predicting welfare than those that targeted *soft* ones. Also expected was the number of pre-processing steps and datasets used having a positive and statistically significant relationship with welfare estimation performance. Even more important, we find that the combination of ML and DL significantly increases predictive power – by as much as 15 percentage points – compared to using either alone. Surprisingly, we find that the spatial resolution of the satellite imagery used is important but not critical to the performance as the relationship is positive but not statistically significant. The finding of no evidence indicating that predictive performance of a statistically significant effect occurs over time was also unexpected. These findings have important implications for future research in this domain. For example, the level of effort and resources devoted to acquiring more expensive, higher-resolution SI will have to be reconsidered given that medium resolutions ones seem to achieve similar results. The increasingly popular approach of combining ML,

DL, and TL, either in a concurrent or iterative manner, might become a standard approach to achieving better results.

# 1. Introduction

There is a need to measure the welfare of people and nations. The world's population should be counted, measured, weighed, and evaluated (Jerven, 2017). Behind these rather raw statements hides a more humanistic perspective where measuring and counting people is in support of humanitarian and developmental efforts, targeting, mapping, and monitoring people at risk of food insecurity, famine, poverty and disease (McBride et al., 2021). The effort is supported by the 17 Sustainable Development Goals along with their 169 targets that were adopted by member states of the United Nations as part of the 2030 Agenda (UN, 2015). To balance the agenda's economic, social, and environmental aspects, more timely, reliable, and appropriate ways of collecting and interpreting information on a broad range of human development outcomes are needed (Head et al., 2017).

A rapidly growing scientific literature seeks to develop new methods and/or augment existing methods for measuring welfare and poverty and monitoring the progress towards the attainment of these SDGs. For close to a decade now, the most interesting approach in this area of research has been the combination of satellite imagery (SI) with machine learning (ML), including deep learning (DL), and transfer learning (TL). These learning algorithms – machine learning, deep learning, and transfer learning, are all, essentially, subsets of the field of artificial intelligence. This review focuses on studies at the intersection between ML/DL/TL, satellite imagery and poverty analysis. The SIML (Satellite Image Machine Learning) methodology combines some of the recent achievements from computer science and object recognition research and applies them to the field of human development research. At a conceptual level, the SIML approach shares many similarities with well-known applications such as learning different object categories from imagery e.g., distinguishing dogs from cats in photographs. For several reasons, it is more complicated to train algorithms to estimate, for example, poverty from imagery. This may be partly because labelled training data is less abundant for human development targets than for everyday objects typically found in image databases such as ImageNetn or AlexNet. Another is that poverty and welfare, and how to measure them are strongly contested issues (Gibson 2016); a fact which is not yet adequately reflected in recent works on SIML. Still, this is a rapidly growing area of scholarship. A recent review of 12 studies evinces that the methodology can predict, for example, the DHS welfare asset index with $R^2$ between .45 and .80 (Burke et al., 2021).

Like in most other fields of study, context matters in this domain. Some forms of data types perform better than others in different contexts. For example, in poor and extremely poor regions, night-time light satellite data have been shown to underperform compared to daytime satellite imagery as the former is noted not to vary significantly in such regions (Yeh et al., 2020). This is particularly true for rural regions where the presence of economic activity, including large farms, does not necessarily mean increasing intensity or even the presence of illumination. At the fundamental level, for a dataset to be useful in this endeavour, there must be sufficient variability in the input variable at different welfare levels and daytime imagery tend to provide that.

The conceptualization and measurement of poverty is a complex endeavour. The increasing success rate of the SIML approach in this domain is influenced by the kind of indicator or measure of poverty that is being targeted. In this regard, it is important to note the different forms and manifestations of poverty or welfare so that in measuring or predicting it, both the so-called 'soft' measures (income, nutrition, food consumption, literacy rates, etc.) and the 'hard' measures (mainly physical assets) are taken into consideration or at least acknowledged. The use of the latter in machine vision processes would be expected to yield higher accuracy than the former in poverty prediction. That is, physical indicators of welfare such as roofing quality and type, infrastructure, and farm sizes, among others, can more effectively be detected and classified from high-resolution SI than the quantity of meat consumed by a household in the last week. This constrains its applicability to certain types of poverty which are place-based and have a physical manifestation, rather than those measures which are transient in character. The present review seeks to introduce this paradigm of satellite imagery/machine learning/poverty analysis to a wider audience in the humanities, the social sciences, and the development community. More specifically, the present paper aims *to review the literature on satellite imagery and machine learning from the perspectives of relevance for measuring human development and more specifically, welfare and poverty measures*.

## 1.1 Shifting the frontiers of poverty analysis

A major strand of poverty geography studies the distributional characteristics – identification of poor areas and impoverished populations (Zhou & Liu, 2022). Spatial identification of poverty is important to the extent that it can reveal the spatial heterogeneity and geographical character of poverty and thus aid the prioritisation of poverty alleviation efforts and resource allocation activities (Erenstein et al., 2010). This is critical given the relative and multidimensional character of poverty. The current standard approaches to quantifying welfare – based on face-to-face household interviews – can deliver detailed estimates of poverty, gender roles, the experience of hunger, and many other important indicators of poverty. Such surveys on poverty often measure incomes, consumption, expenditure, or assets. However, such surveys are expensive, time-consuming, error-prone, and difficult to scale up beyond the community or site level without substantial financial investment. Two of the more well-known survey programmes are the World Bank's flagship household Living Standards Measurement Survey (LSMS) and the USAID's Demographic and Health Survey (DHS). With their roots in the 1980s, they have evolved substantially in terms of coverage, methods, and technology. For example, GPS receivers are nowadays part of the standard equipment when visiting sample villages. While unprecedently rich in indicators, most large-scale surveys, such as these two, tend to suffer from long lags between surveys, limited spatial coverage, and high aggregation levels (nation or region) which impede effective monitoring and evaluation of poverty and welfare indicators at the village and household levels where they are most needed (Burke et al., 2021).

Again, given the multidimensional character of poverty, poverty indices are often constructed from these surveys as a single variable would not suffice as an indicator of poverty. A major determinant in the construction of a poverty index is the kinds of data that are available (Zhou & Liu, 2022). Increasingly, the data requirements, as well as approaches to the measurement of poverty, are becoming more sophisticated as the scale of measurement shifts from national and regional levels to the district, town, household, and even individual levels. Work in recent decades has explored poverty estimation using remote techniques (Blumenstock, 2016). New forms of data are aiding this transition toward finer resolutions though some have met with limited success. Applications based on night-time lights such as Keola et al. (2015) were a step forward. While initial efforts in this domain relied on night-time lights (NTL) data – given the strong correlation between nightlight luminosity and traditional measures of economic growth – the application has seen limited application at finer resolutions and in poorer regions (Blumenstock, 2016; Jean et al., 2016). Other major approaches include the use of high-resolution daytime satellite data (Head et al., 2017; Jean et al., 2016), mobile phone metadata (Aiken, Bedoya, et al., 2021; Blumenstock et al., 2015), internet search history and social media activities (Choi & Varian, 2012; Fatehkia et al., 2020; Llorente et al., 2015) and various combinations of these data sources (Pokhriyal & Jacques, 2017; Steele et al., 2017). These applications have been possible largely as a result of the proliferation of big data as well as newer methods in machine learning to process them.

## 1.2 Application of Artificial Intelligence to poverty analysis

The introduction of machine learning, especially deep learning algorithms, has been instrumental in the application of these new 'big data' sources in poverty studies. An array of ML approaches exists but the most utilized in this domain of research could be classified as feature extraction algorithms or feature selection ones. Feature extraction methods include Principal Component Analysis, Linear Discriminant Analysis, and Singular Value Decomposition. The most popular feature extraction algorithms include Neural Networks, Random Forests, Gradient Boosting, Decision Trees, Naïve Bayes, Gaussian models, Vector Machines, and Linear and Logistic Regressions, among others. These algorithms can also be categorised into *unsupervised* – using information that is not labelled to guide the learning of patterns, similarities, and differences; *supervised* – using labelled data to train the machine to produce correct outcomes; and *semi-supervised* – working with both labelled and unlabelled data so that the former is used to train the algorithm while the latter is used to make predictions. In this domain of research, the outcome could be whether specific locations, areas, households or individuals are poor or not as well as estimate the degree of deprivation. The choice, utility and propriety of each technique are largely informed by the structure of the dataset and the task at hand.

While night-time lights, by themselves, have proven less accurate at differentiating between the poor and the ultra-poor, especially in less-developed regions (Yeh et al., 2020), they have proven useful when used in conjunction with daytime satellite imagery which tends to have higher spatial resolutions. Jean et al. (2016), for instance, successfully trained the deep learning algorithm Convolutional Neural Network (CNN) on detail-rich daytime satellite imagery in which paved roads and metal roofs are visible. In doing so, they developed a technique which shows the relationship between features from daytime satellite imagery and night-time imagery – for the latter, lighted areas are indicative of economic activity. Through this approach, the authors were able to predict indicators of poverty at the regional level.

With regard to mobile phone metadata, there is a strong association between mobile phone use and regional distribution of wealth. Eagle et al. (2010), for instance found that network diversity alone accounted for more than three-quarters of the variance in a region's economic status in the United Kingdom. The application of this approach is not limited to developed countries though, given the increasingly high mobile penetration rates even in developing countries. Blumenstock et al. (2015), for instance, constructed a composite wealth index using principal components of various wealth indicators gleaned from the 2007 and 2010 DHS for Rwanda as well as a phone survey and CDR such as calls and text messages. Through this, the authors demonstrate that a mobile phone subscriber's wealth status can be inferred from their historical phone use pattern, with cross-validated correlation coefficients of 0.68. The authors accomplish this through a combination of feature engineering and feature selection to transform phone users' transaction logs into metrics that are then winnowed through various dimension reduction techniques. The authors also demonstrate how alternative supervised learning models, including decision tree-based regressors and classifiers, can produce comparable results. In addition to call and text messaging history, other key CDR features include top-up patterns, handset type, and user mobility between and among cell towers (Steele et al., 2017). Indeed, merely owning a mobile device is indicative of a certain level of welfare.

Exponential growth in computing power and more effective and efficient learning algorithms are providing conducive conditions for combining disparate data sources to predict and estimate poverty and welfare. Pokhriyal and Jacques (2017), for example, employed a Bayesian Gaussian Process regression on CDR, satellite imagery and environmental data to accurately predict poverty even at the individual level in Senegal. They demonstrate superior prediction accuracy when using such disparate data sources compared to using single datasets. Similarly, Steele et al. (2017) employed multiple data sources – CDR, poverty data, and remote sensing covariates such as night-time lights, vegetation indices, and metrics on distances to roads or urban centres – to predict poverty levels in Bangladesh using hierarchical Bayesian Geostatistical Models.

It is important to note that CDRs are usually not easily accessible from mobile network operators due to commercial rights and privacy concerns. Even with scanty CDR, enough insights can be gleaned when such data is combined with other data sources. Njuguna and McSharry (2017), for instance, demonstrate how such sparce CDR can be combined with per capita mobile handset ownership and call volume per handset, normalized night-time light from satellite imagery, and population density to estimate a multi-dimensional poverty level in Rwanda, with a cross-validated correlation coefficient of 0.88. The accuracy of these studies is not overly negatively impacted by applying them to multiple countries either. For example, using a variety of datasets including the DHS malnutrition and asset poverty data, remotely sensed solar-induced chlorophyll fluorescence, as well as precipitation and conflict data, Browne et al. (2021), demonstrate how random forest models can be used to estimate malnutrition and poverty prevalence across 11 countries – Bangladesh, Ethiopia, Ghana,

Guatemala, Honduras, Kenya, Mali, Nepal, Nigeria, Senegal and Uganda. Here, we hypothesize that while the combination of methods and data sources can improve the accuracy of prediction of welfare, performance reduces as the application is spread over multiple countries and regions. This may be due mainly to the multidimensional nature of poverty and the different ways in which poverty spatially manifests. Thus, *while it is useful to assess how accurately human development can be predicted, it is equally important to evaluate the performance of the different approaches and datasets*.

While the performance of models for poverty analysis in this area of scholarship is reported to be on a general ascendency (Burke et al., 2021), certain factors play a role in the predictive performance. Some important indicators can be measured with higher success than others. Model performance may, thus, be influenced by the type, spatial resolution, and nature of the data used in the modelling. Poverty data, for example, is usually sourced from variables based on income, consumption and/or assets. The predictive power of assets has been shown to be higher than consumption and income-based variables (Jean et al., 2016). This is partly because consumption and income levels tend to vary more significantly within shorter periods of time as they relate more directly to harvest outcomes, job losses and gains, and even household size (Steele et al., 2017). Assets such as ownership of mobile phones, tractors, as well as the type and quality of roofing of buildings are more durable and tend to vary less often over time. Even more importantly, some poverty indicators, such as assets, are more easily discernible in high resolution satellite imagery – the so-called hard indicators – than other indicators such as consumption and income.

Furthermore, despite the increasing access to big data, higher resolution datasets are usually more difficult or expensive to access and computationally more taxing to process. Some studies employ pre-processing steps such as pan-sharpening to improve the spatial resolution of satellite data. For example, night-time lights, which tend to have lower spatial resolution of about 1km, can be fused with higher-resolution daytime satellite imagery to produce a higher resolution composite dataset. Still, machine learning algorithms have been used to infer individual subscribers' socioeconomic status directly from their individual phone use habits (highest resolution) and then aggregate such predictions to town, district and regional levels (lower resolutions) (Blumenstock et al., 2015). We hypothesize that certain types of assets and features can more effectively be measured in higher-resolution imagery than others. This will mean the spatial resolution of the dataset which is used for the analysis becomes important for model performance. Given the trade-off between increasing spatial resolution and accessibility of data, a clearer understanding of the cut-off point *at which ML algorithms most accurately predict welfare* is most relevant for the further development of the SIML approach.

For the wider user community – social science researchers and the development practitioners – concerned with tackling poverty, inequality, and improving welfare, understanding which ML approaches best predict welfare, its accuracy and even at what spatial resolution this can most optimally be done are fundamental questions. Shedding light on these, based on the state-of-the-art in this domain of scholarship, is not only important for future methodological development of the field but also has long-term policy implications. It can, for instance help to understanding why there are currently scanty downstream applications of this approach in development (Burke et al., 2021), with a few recent exceptions (Aiken, Bedoya, et al., 2021; Aiken, Bellue, et al., 2021; Blumenstock et al., 2021).

## 2. Methodology

### 2.1 Inclusion criteria

Our review method can be described as integrative rather than systematic (Snyder, 2019). This body of knowledge, mixing preprints, working papers, technical reports, peer-reviewed papers, and conference papers with contributions of various disciplines is notoriously difficult to capture with one single approach. We used Xie et al. (2015) — one of the first publications to examine the application of ML and SI for measuring poverty and economic well-being — as the benchmark and narrative for our analysis. On this basis, papers completed prior to 2014 and do not apply ML to study socioeconomic wellbeing from SI were excluded from the study. We include literature from published journal articles, grey literature such as working papers and validation studies that have clear empirical application, i.e. we excluded reviews of literature. We however limited our inclusion criteria for the year of publication or completion of the drafts of the grey literature. For study design, we included any study that sufficiently describes the application of AI, ML and DL on SI. On selection criteria based on population and geographical location, we had no restrictions, meaning that studies from high-, middle-, and low-income countries were eligible for inclusion. For thematic focus, we included studies that explicitly describe or propose either conventional or new ways of measuring the welfare or poverty levels of populations or proxies for doing so within social science disciplines.

We gathered papers from multiple sources using different search words, phrases and topics related to the subject of the study. We focused on the use of SI or data, prediction of socioeconomic welfare indicators within the timeframe specified earlier. Since our interest was on both peer-reviewed papers and grey literature, we did not restrict our search to any specific search engines. However, we accessed papers on the former from Google Scholar and ScienceDirect. Our final database from this search comprised 60 papers from peer-reviewed journal articles, preprints, conference presentations, and working papers and other grey literature.

## 2.2 Data preparation, regression model specification and description of variables

### 2.2.1 Data preparation

Some studies report multiple results for the same metric used in analysing the target outcome, making it difficult to record results for such studies as a single variable. The multiple results may be estimated and reported for different datasets, models, study locations, years or a combination of these. To identify the actual performance of models, where separate results are presented for training and validation datasets (e.g. Hofer et al., 2020), we recorded the results for the latter. For results of different models, we recorded the results of the model with the highest precision (e.g. Mahabir et al., 2020). Where separate results are reported for satellite and ground truth data, the former is used (e.g. Bruederle & Hodler, 2018). Where several models are run to observe control effects, the results of the full model are used (e.g. Bruederle & Hodler, 2018).

A number of studies looked at several target outcome variables. Where studies report multiple indicators, we captured the main target welfare outcome of interest specified or inferred by the authors. Unless otherwise specified, we report the composite target where there is one among the targets of a study. These are usually related to indices of poverty, inequality, and related measures or proxies of welfare. And in the absence of such explicit targets as described, we capture data on the target closest or related to the indices of economic wellbeing. The primary targets include asset wealth index (AWI), poverty rates, socioeconomic status, and slum mapping, among others. Others target socioeconomic indicators (which were rarely captured in our study because of our focus on the main indicators) including access to electricity, night-time light (NTL), access to water, access to a toilet, educational attainment, monetary income, body mass index (BMI), and population (see for example, Lee et al., 2021; Steele et al., 2017; Tingzon et al., 2019). These are used as proxies for measuring poverty and inequality and yet are not readily observable from daytime satellite imagery.

We define target as the outcome variable or what each paper tries to estimate or predict. The target indicators are measured at different levels; individual, household, neighbourhood, village, enumeration area, etc. Some authors used multiple indicators for measuring their poverty/welfare outcome variable. In such cases, we extracted the main target outcome reported in such papers for the purpose of our regression model. These were referenced in the title of the study, reported in the abstract or in the main text of the papers.

2.2.2 Regression model specification and description of variables

What determines the predictive power of the various satellite imagery and machine learning (SIML) approaches employed for studying welfare and poverty? Answering this question helps shed light on the relevance for human development and the current methodological complexities and data requirements for predicting welfare using methods other than ground truthing and surveys. We answer the above question using regression analysis. Our dependent variable is the explained proportion of variability in the welfare variables measured using the SIML methods applied by the 60 papers. With our dependent variable measured as a proportion, our interest is in the conditional expectation of the share of variability explained by paper $i$, $y_i$, conditional on a vector of covariates, $\mathbf{x}_i$. We therefore specify the following fractional probit model:

$$E(y_i|\mathbf{x}_i) = \alpha + \boldsymbol{\beta}'\mathbf{x}_i + \varepsilon_i$$

where $y_i \equiv 0 < y_i < 1$, $\alpha$ is the intercept, $\boldsymbol{\beta}$ are the coefficient vector associated with the explanatory variables described in Table 1, and $\varepsilon_i$ is the random error term. The vector $\mathbf{x}_i$ has seven variables, which are as follows.

**Spatial resolution:** The first is spatial resolution of the satellite images. The papers reported satellite spatial resolutions in centimetres, meters and kilometres. A majority (about 67%) of the resolutions were reported in meters and as a result, we converted all resolutions to meters. We expect the predictive power of a study to be positively correlated with spatial resolution[1].

---

[1] We inverted the spatial resolution variable so that high values would intuitively be interpreted as better resolution.

**Number of pre-processing methods:** The second variable in the vector $\mathbf{x}_i$ is the number of pre-processing methods used by the various publications. Making meaning and preparing satellite images for analytical purposes as well as linking them to ground truth data requires additional data transformation steps which are referred to as pre-processing. Depending on the number and complexity of satellite imagery involved, one or more methods of pre-processing the data may be required. Therefore, we expect the proportion of variability explained by the papers to be increasing with the number of pre-processing methods used.

**Number of datasets used:** The vector $\mathbf{x}_i$ also contains the number of datasets used by each paper for predicting welfare. While some of the studies use SI and related datasets, others use a combination of SI methods and ground truth data. We expect the power of the predictions to be increasing as more datasets are used for training.

**Target welfare indicator:** Different welfare indicators were used in the sample of published papers included in this study. These included household-level measures such as poverty and inequality indices, mobile phone use, and expenditure; community and neighborhood-level indicators such as mapping slams, infrastructure quality, and areas lit at night in square kilometers; and city and country-level indicators such as economic development, gross domestic product, and employment rates. Given the sample size and the concentration of indicators, we constructed a binary variable that took on the value one for 'soft' welfare indicators (e.g., income, expenditure, and quantity of proteins consumed) and zero for 'hard' welfare indicators (e.g., physical assets such as type of roof, rail tracks and bridges). We expect that SI and related methods used by the published papers would more accurately predict 'hard' indicators than 'soft' indicators, that is, we expect this variable to be positively correlated with the predictive power of the models estimated by the papers.

**Type of method applied:** Fifth, the vector $\mathbf{x}_i$ captures methods of Artificial Intelligence (AI). The 60 papers apply two AI methods—deep learning and/or machine learning. Our hypothesis is that type of method matters for the predictive power of the models presented in the papers. Therefore, we constructed three dummy variables, which are where the study applied (a) deep learning only, (b) machine learning only, or (c) a combination of deep learning and machine learning. We expect that combining both approaches would increase the predictive power of the models over and above using either one of the approaches.

**Number of countries:** Penultimate, $\mathbf{x}_i$ also contains a binary covariate which captures whether the datasets used for the studies covered one or more countries. We expect that there would be an inverse relationship between the number of countries included and the performance of the model as spatial poverty tend to manifest differently in various countries.

**Year study was published:** Lastly, our regression model includes year of publication as a covariate. As knowledge improves and better approaches and methods become available, one could expect prediction accuracy to rise. Thus, we expect prediction power to increase with time, which is why we included the year of publication, expecting a positive and significant association with the proportion of variability explained by the models.

Table 1 shows the summary statistics of the dependent variable and explanatory variables. The predictive power of the models in the 60 articles included in our analysis ranged between 36% and 96% with a mean of 75% and a median of about 79%. Spatial resolution of satellite images reported in the papers range from just about 5cm to 1000m; the mean and median were about 199m and 3.8m respectively, meaning that the resolutions reported are highly skewed to the right. In fact, if we use 5m/pixel as benchmark for high resolution, we observe that about 53% of the papers report using high resolution imagery. Some studies reported no pre-processing; the mean number of pre-processing methods was about two. On average, the studies used nearly four different datasets for their analysis, and less than half of the studies (about 28%) targeted 'hard' welfare indicators.

**Table 1: Description and summary statistics of study variables**

| Variable | Min | Max | Mean | Std. Dev. |
|---|---|---|---|---|
| Proportion of variability explained by models | 0.36 | 0.96 | 0.75 | 0.16 |
| Spatial resolution of satellite imagery (meters) | 0.05 | 1000 | 198.82 | 2.81 |
| Number of preprocessing methods | 0 | 6 | 2.25 | 1.61 |
| Number of datasets used | 1 | 8 | 3.67 | 1.68 |
| Target welfare indicator studied (0 if soft, 1 if hard) | 0 | 1 | 28.33 | - |
| Proportion using Deep Learning (DL) | 0 | 1 | 0.38 | - |
| Proportion using Machine Learning (ML) | 0 | 1 | 0.50 | - |
| Proportion using both DL and ML | 0 | 1 | 0.12 | - |
| Number of countries studied | 1 | 57 | 5.90 | 11.45 |
| Year of study or publication | 2014 | 2021 | 2018 | 1.81 |

Note: N=60

On type of AI method applied, about half of the studies reported using machine learning and 38% used deep learning, and 12% used both. About 58% of the papers were based on country-specific studies but the mean number of countries covered per study was approximately six. The 60 papers considered in this article were published between 2014 and 2021.

## 3. Regression Results

The regression results are presented in Table 2 and the key insights are as follows. First, the mean spatial resolution of satellite imagery has a positive but statistically insignificant effect on the predictive power of the published welfare analyses. This is a surprising result because our a priori assumption was that SI with a higher spatial resolution would be significantly associated with a better prediction of welfare. After examining several functional forms of the variable, the aggregate results from the pool of studies in our sample do not show a significant effect.

Second, the number of preprocessing methods used has a significant positive effect on predictive power at the 0.10 level. An additional method of preprocessing is associated with almost two percentage points increase in the proportion of variation in welfare explained by the papers included in this article.

Third, the number of datasets used by a study has a positive and highly significant ($p$-value = 0) correlation with the predictive power of the welfare estimates. This typically means that studies that use a combination of ground truth and satellite data have higher welfare prediction power. Specifically, an additional dataset increases the predictive power of the welfare estimate by about four percentage points.

Fourth, the type of welfare indicator targeted by satellite imagery and the AI methods matters. As could be expected, studies targeting 'hard' welfare indicators have higher predictive power than those targeting 'soft' indicators and the difference in predictive power is significant at the 1% level. Compared with 'soft' indicators, targeting 'hard' indicators is associated with about 17 percentage points higher predictive power; this is a large magnitude of difference.

Fifth, we find that choice of tool for a study (whether machine learning, deep learning, or a combination of the two) matters for the predictive power of the welfare models. Using machine learning increases predictive power by about seven percentage points compared with using deep learning but the effect is statistically significant only at the 0.10 level; using a combination of the two increases predictive power by 15 percentage points relative to using deep learning, and this effect is significant at the 1% level. Similarly, Figure 1 shows that using both machine and deep learning increases the proportion of explained variation in welfare indicators by about eight percentage points above what could be realized using machine learning alone ($p$-value = 0.022). This means that combining DL and ML should be preferred to either of them as a single tool in predicting welfare.

Table 2. Determinants of welfare prediction performance

| Variables | (1) Coefficient | (2) Average marginal effects |
|---|---|---|
| Inverse spatial resolution of satellite images | 0.008 | 0.002 |
|  | (0.011) | (0.003) |
| Number of preprocessing methods | 0.052* | 0.016* |
|  | (0.030) | (0.009) |
| Number of datasets used | 0.125*** | 0.038*** |
|  | (0.032) | (0.010) |
| Hard vs. soft target welfare indicator | 0.607*** | 0.167*** |
|  | (0.130) | (0.033) |
| AI method (reference is Deep learning) |  |  |
|    Machine learning | 0.212* | 0.064* |
|  | (0.122) | (0.036) |
|    Machine learning + Deep learning | 0.534*** | 0.138*** |
|  | (0.142) | (0.032) |
| Cross-country vs. country-specific studies | –0.036 | –0.011 |
|  | (0.103) | (0.031) |
| Year of publication | 0.015 | 0.004 |
|  | (0.034) | (0.010) |
| Intercept | –30.071 |  |
|  | (69.005) |  |
| Observations | 60 |  |
| Pseudo R-squared | 0.050 |  |
| Model Chi-squared | 63.587 |  |
| p-value for model test | 0.000 |  |

Note: Robust standard errors in parentheses; *** p<0.01, ** p<0.05, * p<0.1. AME denotes average marginal effect.

Sixth, although we observe, as expected, that the number of countries included in a study is negatively associated with the predictive power of the welfare estimates, the effect is not statistically significant at conventional levels. One reason that may be responsible for this is the lack of standardization of welfare measurements and indicators across countries.

Finally, although we expected a priori that the predictive power of the welfare estimates would have improved over time, we find no evidence of a statistically significant effect even though the expected positive sign is observed.

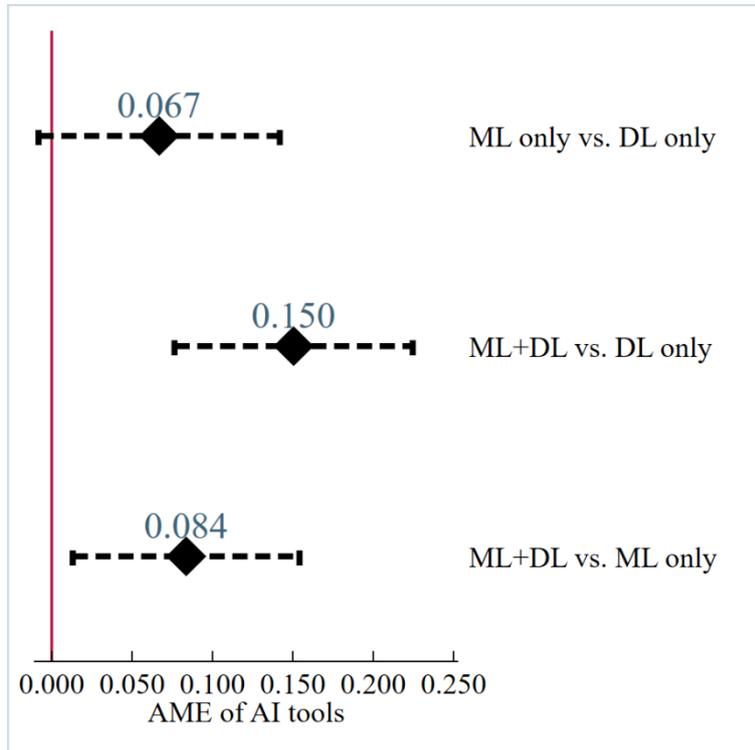

Figure 1. Difference in average marginal effects (AME) of AI tools on predictive power.
Note: The line caps represent 95% confidence intervals.

## 4. Discussions

The present paper set out to review the state-of-the-art at the intersection of the application of cluster of ML and DL tools on satellite imagery to predict poverty or welfare. Key findings from this nascent but rapidly growing field suggests the following. First, our finding that while positive the relationship between the mean spatial resolution of the individual studies and their predictive power is not statistically significant was quite surprising. Conventional wisdom holds that higher-resolution satellite imagery would contain more abundant information about the landscape and its features that could be correlated with economic activity (Jean et al., 2016). It then stands to reason that training datasets based on such higher-resolution imagery would produce more accurate prediction and produce models that have higher predictive power (Engstrom et al., 2021; Head et al., 2017). Our result suggesting a positive but statistically insignificant relationship between spatial resolution and accuracy has important implications. It suggests, for instance, that previous poor results achieved were down to other factors than the unavailability of higher spatial resolution satellite data *per se*. For researchers, this implies that going forward, additional resources would not need to be expended to acquire higher resolution imagery which are often only commercially-available at high cost (Ayush et al., 2020) and that publicly available satellite imagery would suffice in most cases.

While we do not find any evidence of a statistically significant effect that prediction performance using this approach increases over time, Burke et al. (2021) find the contrary. It must be noted, however, that they assessed the performance of these approaches within the specific domains of smallholder agriculture, economic livelihoods, population, and informal settlements. They also attribute the improving performance they measure to three main factors: more creative application of advances in computer visions, more abundant and higher quality satellite imagery, and more numerous and accurate training datasets. The later jives with our findings which suggest that the number of datasets used is positively and statistically significant for prediction performance. In this vein, the increasing proliferation of more accurate and higher quality training datasets portend well for this field of scholarship. Most studies in this area previously relied more heavily on night-time lights datasets with coarse spatial resolutions (1km/pixel) for estimating the level of welfare or development. The cluster of ML approaches recently applied in this intersection have proven to significantly improve predictions that could be achieved using NTL. For example, one of the most important findings from Yeh et al. (2020) was that nighttime lights tend to perform relatively poorly compared to daytime imagery in predicting asset wealth, largely because the former does not vary sufficiently in poor regions. The review also notes the limited downstream application which it attributes, in part, to the novelty of the approaches and their lack of interpretability. With regards to the latter, explainable AI is the next rung in the ladder of applying ML to everyday social development issues such as poverty analysis. This requires transparency in model building (Hall et al., 2022). We argue that a necessary, even if insufficient, condition for the development of transparent, explainable, and interpretable rather than black box (Rudin, 2019) ML models is adequate domain knowledge which, in turn, requires co-option of development researchers and practitioners.

An important consideration in the use of satellite imagery is the type and number of pre-processing operations that are employed to format images before they are fed to the models for training. In the reviewed papers, the main pre-processing operations include radiometric correction, rotation and flipping, channel re-scaling, normalization, cropping, and pan-sharpening. The most-used pre-processing step – pan-sharpening – entails enhancing the lower spatial resolution of multispectral band images by combining them with higher resolution panchromatic images (Hofer et al., 2020). That it is the most employed pre-processing operation is unsurprising given the conventional view that higher spatial resolution datasets invariably translate into better training data and more accurate predictions. However, if it turns out that the relationship between the spatial resolution of the SI and accuracy of the result is positive but statistically insignificant, then pan-sharpening might become a redundant operation. This notwithstanding, other pre-processing steps would remain critical to ensure more accurate prediction of welfare. In the reviewed papers, radiance correction remains key for filtering out ephemeral light sources in DMSP data (Kim et al., 2016), especially in studies that rely on night-time lights (Bruederle & Hodler, 2018; Rybnikova & Portnov, 2020; Zhao et al., 2019). Given how fundamental some of these pre-processing operations are, some SI datasets such as PlanetScope imagery are often radiometrically corrected before delivery to users (Warth et al., 2020), while for others, users need to implement the correction (Duque et al., 2017; Leonita et al., 2018).

Key take aways from our results is that the total number of datasets used in training the of the model, the nature of the target welfare indicator and the specific learning model contribute the most to explaining the level of welfare that can be predicted. With regards to capability of the various learning models, our results suggest that a combination of more conventional ML models and those using DL approaches has the most predictive power in welfare estimation studies using SI data. A vast majority of the reviewed studies employed either ML or DL alone (24 and 30, respectively), with only 7 of them (Gram-Hansen et al., 2019; Hofer et al., 2020; Lee & Braithwaite, 2020; Li et al., 2019; Mboga et al., 2017; Puttanapong et al., 2022) combining both ML and DL in their analyses. These papers combine the approaches differently though. While Li et al. (2019) compare their performance concurrently to determine which performs relatively better, Lee and Braithwaite (2020) combine them in an iterative manner. The latter use the ML algorithm of eXtreme Gradient Boosting (XGBoost) to estimate welfare levels for all populated places in 25 countries and then use this predicted welfare level to train the DL model of convolutional neural network (CNN). This was done to circumvent the need to use night-time luminosity. Lee and Braithwaite (2020) then fed the featurized information which was the output from the CNN model back to the XGBoost model. They then apply transfer learning from the second iteration onwards to augment learning and speed up the process. Combining these different models in the iterative manner through transfer learning tends to generate better training data which contribute to prediction accuracy (Burke et al., 2021; Head et al., 2017; Hofer et al., 2020).

Our finding of significantly higher predictive power of models which are based on visible features, the so-called 'hard' indicators, is instructive even if unsurprising given that 'soft' indicators such as income levels, expenditure or the quantity of meat consumed by a household, for example, are more difficult to estimate from an SI than the existence and size of buildings or quality of roofing in a scene. In this sense, these models may be grouped into feature-based algorithms – those that rely on quantifiable geospatial features such as the number of building, length of road, number of junctions, etc and image-based models – those that can recognize the qualitative characteristics of these features (Lee & Braithwaite, 2020). The choice of either of these then comes down to the resolution of the satellite data available since lower resolution SI tend to provide more information about the spatial context such as whether the data is from a rural or urban landscape while higher resolution SI are more useful for extracting the qualitative characteristics of the features (Kim et al., 2016). This suggests that the increasingly more accurate results obtained by studies in this area of scholarship could be driven more by a combination of improving the spatial resolution of readily available SI data and, perhaps more importantly, the potency and effectiveness of new tools, and approaches as well as the computational power to implement these as Burke et al. (2021) contend. As an illustration, Lee et al. (2021) show the importance of 'hard' indicators such as infrastructure (rail tracks and bridges) as well as other physical features like vehicles, street lights and billboards as indicators of the existence of services and development, and by extension, welfare. It is little wonder that the recent trend is one of combining feature-based and image-based approaches in either sequential or complementary manner to predict welfare (Abitbol & Karsai, 2020; Lee & Braithwaite, 2020; Warth et al., 2020).

## 5. Conclusions

The paper set out to review the state-of-the-art of studies that fall at the intersection of satellite imagery (operationalised broadly as all remotely sensed data), poverty analysis and using artificial intelligence tools of machine, deep, and transfer learning. Studies in this nascent domain of scholarship is reported to be seeing consistently improving accuracy over the last couple of years, though our results suggest a weak but positive relationship between the spatial resolution of the SI in use and prediction accuracy. The strong explanatory power of models that are based on 'hard' indicators, which are, in turn, more accurate on higher-resolution SI data suggests that marked improvements in the tools and computational capabilities are instrumental in these continuous improvements in prediction accuracy in the last few years. It also points to the importance of intermediate data pre-processing steps, especially those related to improving SI resolution such as pan-sharpening. Further progress in algorithms as well as less expensive access to more accurate and numerous training datasets portend well for the field of application of welfare measurement using SI and machine learning tools. With regards to the specific models and their efficacy in predicting poverty and welfare, we found that a combination of ML and DL algorithms have the best performance, compared to either individual group of models in their own right. This further supports our view that rather than the sheer improvements in the resolution of SI, it is the increasing efficacy of newer tools and models that could be spurring any improving results that we may seeing in this area.

While the application of the SIML approach continues to see improving model performance, more transparency is needed to achieve the next target – explainable AI. This is quite a daunting task given the multidimensional nature, place-based character of poverty. Thus, the differences in spatial manifestation of poverty and its markers in different locations and regions or countries, from a ML model perspective, further complicate this approach. It is mainly for this reason that the combination of multiple datasets or data sources enhances the performance of these models.

**Acknowledgements**

The authors would like to thank financial support during the project from the Swedish Research Council 2019-04253 and Riksbankens Jubileumsfond MXM19-1104:1.